\documentclass{article}

\usepackage{arxiv}
\usepackage{graphicx,natbib,amssymb,amsmath,bm}
\usepackage[utf8]{inputenc} 
\usepackage[T1]{fontenc}    
\usepackage{hyperref}       
\usepackage{url}            
\usepackage{booktabs}       
\usepackage{amsfonts}       
\usepackage{nicefrac}       
\usepackage{microtype}      
\usepackage{lipsum}		
\usepackage{graphicx}
\usepackage{natbib}
\usepackage{doi}
\usepackage{subeqnarray}

\title{Multi-step dual control for exploration and exploitation in autonomous search with convergence guarantee}

\author{{\hspace{1mm}Yuan~Tan} \\
		The School of Automation\\
	Southeast University\\
	Nanjing, China \\
    \texttt{tan\_yuan@aliyun.com} \\
	\And
	{\hspace{1mm}Jun~Yang} \thanks{Corresponding author}\\
	Department of Aeronautical and Automotive Engineering\\
	Loughborough University\\
	Loughborough, LE11 3TU, UK \\
	\texttt{j.yang3@lboro.ac.uk} \\
		\And
	{\hspace{1mm}Wen-Hua~Chen} \\
	Department of Aeronautical and Automotive Engineering\\
	Loughborough University\\
	Loughborough, LE11 3TU, UK \\
	\texttt{W.Chen@lboro.ac.uk} \\
	\And
		{\hspace{1mm}Shihua~Li} \\
		The School of Automation\\
	Southeast University\\
	Nanjing, China \\
	\texttt{lsh@seu.edu.cn} \\
	}

\renewcommand{\shorttitle}{\textit{arXiv} Template}

\hypersetup{
pdftitle={Multi-step dual control for exploration and exploitation in autonomous search with convergence guarantee},
pdfauthor={Yuan~Tan, Jun Yang, Wen-Hua Chen, Shihua Li},
pdfkeywords={Autonomous Search, Multi-step dual control, Recursive feasibility and convergence, Path planning, Exploration and exploitation.},
}

\begin{document}
\maketitle

\begin{abstract}
Motivated by the recently proposed dual control for exploration and exploitation (DCEE) concept,  this paper presents a  Multi-Step DCEE (MS-DCEE) framework with guaranteed convergence for autonomous search of a source of airborne dispersion.  Different from the existing stochastic model predictive control (SMPC) algorithm and informative path planning (IPP) approaches, the proposed MS-DCEE approach uses the current and future input to not only drive the agent towards the estimated source location (exploitation) but also reduce its estimation uncertainty (exploration) by actively learning the operational environment. Unknown source target position, together with unknown environment, impose significant challenges in establishing the recursive feasibility and the convergence of the proposed algorithm. To address them, with the help of the property of Bayesian estimation, we develop a two-step approach where the unbiasedness of the mean estimation is assumed first and then the randomness of the mean estimate under each collected information sequence is accounted. Based on that, we develop a MS-DCEE scheme with suitable terminal ingredients where recursive feasibility and convergence are guaranteed. Two simulation scenarios are conducted, which show that the proposed MS-DCEE algorithm outperforms the SMPC, the IPP and the single-step DCEE approaches in terms of searching successful rates and efficiency.
\end{abstract}


\section{Introduction}
In recent years, the events involving the release of hazardous gases have frequently occurred, which becomes a key area of concern for industry and the public \citep{TTYY2,TTYY1}. Detecting and then locating the source of harmful gas leakage in a timely manner is of great significance to enable subsequently proper handling of threats and accidents \citep{Y1,Y2}. It is also important for environment protection where law enforcement needs to detect and find the sources of deliberately, accidentally or naturally released harmful gases (e.g., large quantity of methane emission  from an unused oil field or an animal farm). Equally search through airborne dispersion also widely exists in nature, from moths seeking mating with pheromone to polar bears locating food using odour.

Mobile sensor platforms have been recently emerging as a promising new technology in searching the source of airborne dispersion where a mobile ground robot or an unmanned aerial vehicle (UAV) is equipped with chemical or biological gas sensors to detect and seek the source. It provides a much more flexible and cost effective approach in autonomous search, so attracting a great deal of interest recently \citep{Burgu}. Autonomous search algorithms for source localization with a mobile robot can be roughly classified into two categories: informative path planning approaches and control oriented algorithms.

Control-oriented search is to seek a source from a control engineering perspective in a broad sense. In this category, there are a number of approaches. Inspired by biology, chemotaxis \citep{che} and anemotaxis \citep{XX} search sources using chemical gradients and others as  cue. By exploiting the fact that the maximum chemical concentration is located in the source location, extremum seeking has been adopted and developed into autonomous search where perturbation signal is added into the search action \citep{Azzollini,Liu2010}. A promising feature of the control driven search is that certain theoretic properties such as stability and convergence can be established under certain condition.  As indicated in \cite{C1}, it is possible to formulate an autonomous problem as a Stochastic Model Predictive Control (SMPC)  problem by defining the cost  based on the expected error between the robot's future position and the current estimation of the source location. Therefore SMPC can be exploited to plan the path for  the autonomous search in a receding horizon fashion \citep{34}.

Informative path planning (IPP) approach considers the search process as an information gain process about the source position and local environment so as to plan an optimal path in terms a specified information measure \citep{Y1,Y2,Y7,Y9,TTTYY1,TTTYY2,TTTYY3}, which can be thought of as a strategy to explore the environment.  Infotaxis \citep{Y1} has been proposed and implemented in indoor experiment to estimate a source location using a ground mobile robot. With entropy as an information measure, Entrotaxis has been proposed in \cite{Y2}, and first implemented in an indoor environment and then extended to source search in outdoor environment with a UAV.  More recently,  \cite{TTTYY2} extended the Entrotaxis based planning approach as Entrotaxis-jump algorithm to search source in  large-scale road networks. To solve the random blocking source search problem, an IPP approach combining Infotaxis and Entrotaxis is proposed in \cite{TTTYY3}. Different from control oriented approaches, so far there is no theoretical tool to analyze the convergence of IPP search strategy.

Reduction of the environment uncertainty through the sensing process is often dependent on the location and trajectory of the robot agent or control input. Therefore, combining the advantages of SMPC and IPP, a new control framework of DCEE is recently proposed in \cite{C1} by exploiting the dual control concept.
The cost function of DCEE, different from the SMPC, is defined as the expected error between the agent's predicted future position and the predicted future estimate of the source by using predicted \emph{future measurements}. In the DCEE framework, it is clear that the control input is determined by exploiting its dual effect: i.e., exploitation of the current information by driving the agent towards the predicted estimated source location (somehow as in SMPC)  and exploration of an unknown environment to reduce source parameter uncertainties as in IPP.

While exhibiting a promising performance, the DCEE framework lacks of rigorous performance analysis such as recursive feasibility and convergence of DCEE that are of great importance \citep{C1}.
SMPC considers the problem of tracking a well-defined reference or set point. The recursive feasibility and convergence can be ensured by including a terminal set and terminal cost in the optimization problem \citep{34}. However, the control goal of the MS-DCEE in autonomous search is to seek a target at an  \emph{unknown} location within an \emph{unknown} environment. Environment factors such as wind speed and direction affect the dispersion greatly. If the aim of control is to drive the robot close to an estimated source location, it continuously changes with the update of the source estimation, which is only available in the future after new data arrive. This presents significant difficulties for recursive feasibility and convergence analysis. Furthermore, there is a strong coupling between the uncertainty quantification (e.g. using Bayesian inference)    and planning which makes the analysis difficult.

The main contribution of this paper is to develop a DCEE algorithm for autonomous search with proven properties including recursive feasibility and the convergence. In our setting, Bayesian filtering is utilized to provide the preview of future uncertainty information and the predictive posterior mean and covariance of the unknown source are exploited to reformulate the multi-step cost function. With the help of the properties of Bayesian estimation, a two-step approach is proposed to tackle the challenges arising due to the unknown target position and unknown environment in DCEE. Furthermore, it is also realised that although the mean given by Bayesian estimation is unbiased statistically, at each run of the autonomous search, the mean yielded by Bayesian estimation may be biased and changes with the collected data sequence. We exploit the fact the mean-square error of the Bayesian estimation is bounded.  First we consider the baseline case where the mean estimation yielded by Bayesian inference is unbiased and establish the recursive feasibility and provide the convergence properties. Then we take into account the randomness of the mean estimation at each run with the help of the boundedness of the mean-square error. This paper also extends the single-step DCEE in \cite{C1} to Multi-Step DCEE (MS-DCEE). Since DCEE could be considered as the combination of IPP and SMPC approaches to some extent, the work in this paper will also help to develop proven properties for autonomous source search strategies developed by information theoretic approaches such as IPP \citep{Y1,Y7,TTYY1}.

{\emph{Notation}:} For the sets $A\subseteq \mathbb{R}^n$ and $B\subseteq \mathbb{R}^n$, $A\oplus B=\{a+b|a\in A,b\in B\}$, while $a\oplus A$ denotes $\{a\}\oplus A$. The image of the set $X$ under real matrix $M$ is denoted by $M\circ X=\{Mx|x\in X\}.$ The set of non-negative integers is denoted by $\mathbb{N}=\{k|k=0,1,2,\cdots\}$. For a vector $x\in\mathbb{R}^n$, the 2 norm $||x||_Q^2$ represents $x^TQx$. For the sets $C\subseteq \mathbb{R}^n$ and $D\subseteq \mathbb{R}^n$, $x\in C\backslash D=\{x|x\in C,x\notin D\}$. The matrix $I_n$ is an identity one with dimension of $n$.

\section{Problem formulation and motivation}

\subsection{Autonomous search problem formulation}
 Consider the problem of autonomously searching an unknown source location in an unknown environment \citep{C1}. The dynamics system involve a linear time-invariant model of the autonomous agent and a nonlinear model representing sensor behaviour
\begin{eqnarray}
&&\ \ \ \ \ \ \ \ \ \ \ \ \ \ \ \ x_{t+1}=x_t+u_t\label{T1a}\\
&&\ \ \ \ \ \ \ \ \ \ \ \ \ \ \ \ z_t=M(x_t,p^s)+v_t\label{T1b}
\end{eqnarray}
where  $x_t =[p^{x_t},p^{y_t},p^{z_t}]\in \mathbb{R}^3$ is the position of the autonomous agent, and $u_t\in \mathbb{R}^3$ is the control action at current time $t\in \mathbb{N}$.  Eq. (\ref{T1b}) represents the gas concentration measurement equation of the sensor equipped in the autonomous agent, where $z_t\in \mathbb{R}$ is  the sensor reading, $M(x_t,p^s)$ is the true concentration with the release source at $p^s=[p^{s,x},p^{s,y},p^{s,z}]\in \mathbb{R}^3$, and $v_t\in \mathbb{R}$ an additive Gaussian uncertainty imposed on the sensor readings.

 The Gaussian plume dispersion equation \citep{3} is selected as the forward Atmospheric Transport and Dispersion (ATD) model to predict the expected concentration $M(x_t,p^s)$ :
\begin{eqnarray}\label{T303}
M(x_t,p^s)=\frac{q_s}{4\pi\zeta_1||x_t-p^s||}\text{exp}\left[\frac{-||x_t-p^s||}{\zeta}\right]\text{exp}\left[\frac{(p^{s,x}-p^{x_t})u_s\text{cos}\phi_s}{2\zeta_1}\right]\text{exp}\left[\frac{(p^{s,y}-p^{y_t})u_s\text{sin}\phi_s}{2\zeta_2}\right]
\end{eqnarray}
where $q_s$ is the emission rate of the airborne hazardous material, and the meteorological parameters $\phi_s, u_s, \zeta_1, \zeta_2, \zeta$ are the wind direction, the mean wind speed, the diffusivity, the particle lifetime and the mixing coefficient $\zeta=\sqrt{\frac{\zeta_1\zeta_2}{1+u_s^2\zeta_2/(4\zeta_1)}}$, respectively.

Since the agent's position is within the domain of searching space, and the maximum movement of the agent is constrained, we impose the state and input constraint on
\begin{equation}\label{T2}
~~~~~~~~~~~~x_{t}\in \mathcal{X}\subset \mathbb{R}^3,u_{t}\in \mathcal{U}\subset \mathbb{R}^3
\end{equation}
where $\mathcal{X}$ is a convex set of position constraint, and $\mathcal{U}$ is a closed set of admissible control action.

The control objective in autonomous search is to design a controller $u_t$ such that the agent's position $x_t$ finally arrive at the true source location $p^s$. Consequently, the finite-horizon performance cost in model predictive control (MPC) framework \citep{Y101} is formulated as follows:
\begin{eqnarray}\label{TY212}
V(x_t,\mathbf{u}_{t})=\sum_{k=0}^{N-1}\left[||x_{t+k|t}-p^s||_Q^2+||u_{t+k|t}||_R^2\right] +||x_{t+N|t}-p^s||_{S}^2
\end{eqnarray}
 where $\mathbf{u}_{t}=\left[u_{t|t}, u_{t+1|t}, \cdots, u_{t+N-1|t}\right]^T$ is a sequence of current and future control, and $Q\succ0$, $R\succ 0$ are weights applied to the state, and control input, respectively, and $S$ is the terminal weight.
MPC is a well established control method for tracking of the known reference signals, the theoretical properties, such as recursive feasibility and stability can be ensured under certain conditions. However, in the autonomous search problem defined in (\ref{TY212}), the source position $p^s$ is unknown.

\subsection{Stochastic model predictive control}

Since $p^s$ is unknown, the cost function (\ref{TY212}) is not available to derive the control law. To cope with this challenge, the unknown source location can be estimated by Bayesian inference methods according to the measurement information at each step,
which provides a mechanism to develop a SMPC algorithm for autonomous search.

Define $\hat{p}_{t}^s$ as the estimate of the true source location $p^s$ subject to a probability distribution function $\mathbb{P}(\hat{p}_t^s)$. Let $I_t$ represent the vector of all past actions and measurements collected up to and including the time step $t$, that is
\begin{equation}\label{T3}
I_t=\left[z_0,u_0,z_1,u_1,\cdots,z_{t-1},u_{t-1},z_t\right].\nonumber
\end{equation}
Define $\mathbb{P}\left(\hat{p}_t^s|I_{t-1}\right)$ and $\mathbb{P}\left(\hat{p}_t^s|I_{t}\right)$ as  the prior and  posterior distribution of $\hat{p}_t^s$ at
time $t$. Essentially, the posterior distribution is computed by  Bayes' theorem \citep{311,312}
\begin{equation}\label{TT11}
\ \ \ \ \ \ \ \mathbb{P}\left(\hat{p}_{t}^s|I_{t}\right)\propto \mathbb{P}\left(\hat{p}_{t}^s|I_{t-1}\right)\mathbb{P}\left(z_{t}|\hat{p}_{t}^s\right)
\end{equation}
where $\mathbb{P}\left(z_{t}|\hat{p}_{t}^s\right)$ is a Gaussian likelihood function. The mean vector and the covariance matrix of $\hat{p}_{t}^s$ are given by $\bar{p}_{t}^s=\mathbb{E}\left[\hat{p}_{t}^s|I_t\right]$, $P_{t}=\mathbb{E}\left[(\hat{p}_t^s-\bar{p}_{t}^s)(\hat{p}_t^s-\bar{p}_{t}^s)^T|I_t\right].$

To drive the agent towards the estimated source location, the cost function  in SMPC framework \citep{TY11,TY22,Y11} based on the information available is formulated as:
\begin{eqnarray}\label{SMPC}
J_{s}(x_t,\mathbf{u}_{t})=\mathbb{E}\left\{\sum_{k=0}^{N-1}\left[||x_{t+k}-\hat{p}_{t}^s||_Q^2| I_t+||u_{t+k}||_R^2\right]+||x_{t+N}-\hat{p}_{t}^s||_{S}^2\big|I_t\right\}.
\end{eqnarray}
Defining $\tilde{p}_{t}^s=\hat{p}_{t}^s-\bar{p}_{t}^s$, note that $\mathbb{E}[(\tilde{p}_{t}^s)^T|I_t]=0$, $\mathbb{E}\left[||\tilde{p}_{t}^s||_Q^2|I_t\right]=\text{trace}(QP_{t})$, $\mathbb{E}\left[||\tilde{p}_{t}^s||_S^2|I_t\right]=\text{trace}(SP_{t})$. Consequently, Eq. (\ref{SMPC}) is equivalent to 
\begin{eqnarray}\label{SMPC1}
&&\!\!\!\!\!\!\!\! J_s(x_t,\mathbf{u}_{t})=\sum_{k=0}^{N-1}\left(||x_{t+k}-\bar{p}_{t}^s||_Q^2 +||u_{t+k}||_R^2\right) \nonumber\\
&&\!\!\!\!\!\!\!\!  +||x_{t+N}-\bar{p}_{t}^s||_{S}^2+N\cdot\text{trace}(QP_{t})+
\text{trace}(SP_{t}).
\end{eqnarray}
 It can be observed from  Eq.(\ref{SMPC1}) that $\mathbf{u}_t$ doesn't affect $\bar{p}_{t}^s$ and $P_{t}$. In the above SMPC setting, the source located is estimated by a Bayesian filtering where  the measurements up to the current time instant $t$ is used. There is no consideration of the influence of the future control on the estimation accuracy and the uncertainty. This motivates the development of MS-DCEE with actively learning the source and environment to improve the performance of autonomous search in the next section.

\section{Multi-Step Dual Control for Exploration and Exploitation}
In this section, we will develop the main results on MS-DCEE to deal with the autonomous search problem with unknown source location.

Define $\left[\hat{z}_{t+1},\hat{ z}_{t+2}, \cdots, \hat{z}_{t+N}\right]$ as the predicted sensor measurements, and extending the definition of $I_t$ to include future decisions and associated predicted measurements with the horizon gives
\begin{equation}
I_{t+k}=\left[I_t,u_{t},\hat{z}_{t+1},\cdots,u_{t+k-1},\hat{z}_{t+k}\right],k=1,2,\cdots,N.\nonumber
\end{equation}

The predicted posterior distribution of the estimated source location at time $t+k$ is updated by
 \begin{equation}\label{T11a}
\mathbb{P}\left(\hat{p}_{t+k}^s|I_{t+k}\right)\propto \mathbb{P}\left(\hat{p}_{t+k}^s|I_{t+k-1}\right)\mathbb{P}\left(\hat{z}_{t+k}|\hat{p}_{t+k}^s\right).
\end{equation}
As a result in (\ref{T1a}-\ref{T1b}) and (\ref{T11a}), the future measurement $\hat{z}_{t+k}$ will be affected by the control input $u_{t+k-1}$,  which  will further affect the belief of the estimated source location.

Inspired by the above discussion, the objective function  developed in
this paper considers the current and future control ${\bf u}_t$ as decision variables to deliver a promising dual effect to balance position control error by exploiting current estimates of the unknown source location and the associated estimation error by active learning and/or exploring the unknown environment.  Therefore, the cost function of multi-step DCEE is formulated as
\begin{eqnarray}\label{DCEE}
J(x_t,\mathbf{u}_{t})=\mathbb{E}\left\{\sum_{k=0}^{N-1}\left(||x_{t+k}-\hat{p}_{t+k}^s||_Q^2\big|I_{t+k}+||u_{t+k}||_R^2\right)+||x_{t+N}-\hat{p}_{t+N}^s||_{S}^2\big|I_{t+N}\right\}.
\end{eqnarray}
It shall be highlighted that SMPC with active learning attracts a significant interest, see \citep{Y12}. However, the dual control effect are considered on the dynamic system (e.g., agent as in this paper) including state and unknown parameter estimation \citep{dual}. Instead DCEE focuses on dual control effect in estimating unknown target parameters or exploring unknown environment. For more discussion, please refer to \citep{C1}. Furthermore, there are two difference between the cost function in this paper and in \cite{C1}: one is a terminal cost is included in Eq.(\ref{DCEE}). The other is a multi-step cost function is used to improve the performance as will be confirmed in a case study. It shall be mentioned that other terminal elements such as terminal constraints will also be added in the algorithm later to ensure feasibility and convergence.

 Define $\tilde{p}_{t+k}^s=\hat{p}_{t+k}^s-\bar{p}_{t+k}^s$,   and the associated predicted mean and covariance matrix are given by $\bar{p}_{t+k}^s=\mathbb{E}\left[\hat{p}_{t+k}^s|I_{t+k}\right]$ and $P_{t+k}=\mathbb{E}\left[(\tilde{p}_{t+k}^s)(\tilde{p}_{t+k}^s)^T|I_{t+k}\right]$, respectively. For the sake of simplicity,  conditional upon $I_{t+k}$ is neglected in notation in the following derivation. Consequently,  Eq.(\ref{DCEE}) is  reformulated as  the sum of the following two components
\begin{equation}\label{TY23}
J(x_t,\mathbf{u}_{t})= J_{ET}(x_t,\mathbf{u}_t)+J_{ER}(P_t,\mathbf{u}_{t})
\end{equation}
with 
$J_{ET}(x_t,\mathbf{u}_{t})=\sum_{k=0}^{N-1}\left(||x_{t+k}-\bar{p}_{t+k}^s||_Q^2+||u_{t+k|t}||_R^2\right)$ $+||x_{t+N}-\bar{p}_{t+N}^s||_{S}^2$
and
$J_{ER}(P_t,\mathbf{u}_{t})=\sum_{k=0}^{N-1}\text{\text{trace}}(QP_{t+k})$ $+\text{trace}(SP_{t+N}).$ 

It can be observed from Eq.(\ref{TY23}) that the cost function is separated into  the sum of two parts: the first one $J_{ET}$ accounts for moving the agent to the mean of the predicted source location, which is related to exploitation, whereas the second term $J_{ER}$ concerns exploration
 by using control to reduce the covariances of the predicted estimation under hypothesised measurements with a control sequence $\mathbf{u}_{t}$. Within the framework of the proposed MS-DCEE, the control input $\mathbf{u}_{t}$  not only affects the predicted location of the agent, but also affects its predicted measurements, which in turn affect the belief of the estimated source position. In other words, the control input $\mathbf{u}_{t}$ not only drives the agent to the estimated source location but also move the agent towards a position where more information could be collected so leads to smaller perception errors.

The autonomous search is formulated as the following constrained optimization problem in the MS-DCEE approach
\begin{eqnarray}
&&\min_{\mathbf{u}_{t}}J(x_t,\mathbf{u}_{t}) \label{To1}\\
s.t.~&& x_{t+k+1}=x_{t+k}+u_{t+k},\\
&&x_{t+k}\in \mathcal{X},k=0,\cdots,N-1,\label{O2}\\
&&u_{t+k}\in \mathcal{U},k=0,\cdots,N-1,\label{O3}\\
&&x_{t+N}-\bar{p}_{t+N}^s\in \mathcal{T}, x_{t}=x_0\label{To4}
\end{eqnarray}
where $x_{t+N}\in \{\bar{p}_{t+N}^s\}\oplus\mathcal{T} \subseteq\mathcal{X} $ with $\mathcal{T}$ is a terminal constraint.

To derive the recursive feasibility and convergence of the proposed MS-DCEE, we need the following assumptions.

{\bf Assumption 1}. The mean vector $\bar{p}_{t}^s$ of the source parameter estimate is bounded and locates within the search area.  The difference between the mean at the current moment and at the next moment $\delta_t=\bar{p}_{t+1}^s-\bar{p}_{t}^s$ of the Bayesian estimator are bounded. That is, there exist two compact sets $\mathcal{O}$ and $\mathcal{G}$ satisfying
\begin{align}
\bar{p}_{t}^s\in \mathcal{O}\subseteq \mathcal{X}, \delta_t\in \mathcal{G}, \forall t\in \mathbb{N}.
\end{align}

{\bf Assumption 2}. The terminal weight matrix $S$ is chosen as the solution of the following Lyapunov equation
\begin{equation}\label{Y1}
(I_3+K)^TS(I_3+K)-S=-Q-K^TRK
\end{equation}
where $K$ is the control gain to be computed.

In Assumption 1, it is easy to realize that the estimated mean value is in the search space. If the value is not in the search space, it can be appropriately converted to a boundary value.
Based on assumption 1 and assumption 2, the terminal set $\mathcal{T}$ in (\ref{To4}) is chosen such that $\mathcal{T}\oplus\mathcal{O}\subseteq \mathcal{X}$, the search area $\mathcal{X}$ could be adapted for expansion,
and the selected feedback gain $K$ satisfies $(I_3+K)\odot \mathcal{T}\subseteq \mathcal{T}$ or $|\lambda_i(I_3+K)| < 1$, where $\lambda_{i}(I_3+K)$ is $i$th eigenvalue of $I_3+K$ and  $K\circ \mathcal{T}\oplus K_f\circ\mathcal{G}\subseteq \mathcal{U}$, where $K_f=I_3$ is a feedback gain of $\delta_t$.

The solutions to this optimization problem (\ref{To1})-(\ref{To4}) are the value function $J^*(t)$ and the optimal state and input trajectories $\mathbf{x}_{t}^*$, $\mathbf{u}_{t}^*$.
In closed-loop operation, we apply the first element of the optimized input trajectory to the system, leading to the closed-loop system $x_{t+1}=x_t+u_{t}^*$.
\section{Recursive feasibility and convergence}
At present, there is no theoretical performance analysis of any IPP based autonomous search methods (e.g. infotaxis or entrotaxis). Analysis of MS-DCEE is even more complicated due to the involvement of the dual control effect. Simply putting, within the DCEE framework, an information driven IPP strategy is combined with a SMPC strategy to offer a better autonomous search strategy. On the other side, by reformulating the autonomous search as a control problem in the DCEE framework, we are able to get access to rich knowledge and tools to analyze the recursive feasibility and convergence. This is another main motivation for formulating the autonomous search problem in this way. In this section, the asymptotically unbiased estimation properties of Bayesian estimator are presented first. Based on that, the recursive feasibility and convergence of the proposed MS-DCEE controller are then established.

\subsection{Boundedness and asymptotically unbiased estimation}

Although it is well known that the Bayesian estimation is unbiased statistically, this does not imply that at each run of the autonomous search, the source and environment parameter estimation is unbiased. Instead this section, we will establish the boundedness of the mean-square-error which plays a key role in developing a MS-DCEE algorithm with guaranteed recursive feasibility and convergence.

The following lemma can be derived from Theorem 3 in \cite{261} and Theorem 1 in \cite{316} straightforwardly.

{\bf lemma 3}. If $T_n=T_n(X_1,X_2,\cdots,X_n)$ is a point Bayesian estimator of  $\theta$ based on n-samples $X_1,X_2,\cdots,$ $X_n$,   there exist a matrix \  $W$ such that\\
(i). $\mathbb{E}(||T_n-\theta||^2)\leq \text{trace}(W)$;
\\ (ii). $\lim_{n\rightarrow \infty}||\mathbb{E}(T_n)-\theta||^2\rightarrow 0$.

Based on the result of Lemma 3, the bounded and asymptotically unbiased estimation properties of the Bayesian estimator are presented by the following theorem.

{\bf Theorem 4}. For a collected information sequence $I_t$, the mean-square-error of the unknown source location $p^s$ yielded by Bayesian estimation is bounded, and there exists a matrix  $W$ such that
\begin{equation}
~~~~~~~~\mathbb{E}(||\hat{p}_{t}^s-p^s||^2)\leq \text{trace}(W).
\end{equation}
Furthermore, one has
\begin{eqnarray}
&&\lim_{t\rightarrow \infty}\mathbb{E}(\hat{p}_t^s)=\lim_{t\rightarrow \infty}\bar{p}_t^s=p^s,\\
&&\lim_{t\rightarrow \infty}\mathbb{E}(||\hat{p}_{t}^s-p^s||^2)=\lim_{t\rightarrow \infty}\mathbb{E}(||\hat{p}_{t}^s-\bar{p}_{t}^s||^2)=\lim_{t\rightarrow \infty}\text{trace}(P_{t}).
\end{eqnarray}

{\sc Proof:} See Appendix A.

According to Theorem 4, we show that the mean of $\hat{p}_t^s$ converges to the true
source parameter $p^s$ with a bounded mean-square-error. Theorem 4 usually holds in practical implementations of environment estimation \citep{Bonzanini}.

\subsection{Recursive feasibility and convergence of auxiliary MS-DCEE}
The mean vector of the source parameter estimate may have changed in every sampling time, which brings challenges to feasibility and convergence analysis. Consequently, a two-step approach is taken to address that: first we analyze the recursive feasibility and convergence for the baseline case where the mean of the estimated source position is accurate and equal to the true source position, i.e., $\mathbb{E}(\hat{p}_{t}^s)=\bar{p}_{t}^s=p^s$ for $\forall \ t \in \mathbb{N}$ and then we investigate the influence of randomness in estimation.

With this assumption in mind, the optimization problem of the auxiliary MS-DCEE with a terminal cost and a terminal set reduces to
\begin{eqnarray}
&&\min_{\mathbf{u}_{t}}\tilde{J}(x_t,\mathbf{u}_{t}) \label{TO11}\\
s.t.&& x_{t+k+1}=x_{t+k}+u_{t+k},\\
&&x_{t+k}\in \mathcal{X},k=0,\cdots,N-1,\label{O21}\\
&&u_{t+k}\in \mathcal{U},k=0,\cdots,N-1,\label{O31}\\
&&x_{t+N}-p^s\in \tilde{\mathcal{T}},x_{t}=x_0\label{TO41}
\end{eqnarray}
where $\tilde{\mathcal{T}}$ is a terminal set, and
$\tilde{J}(x_t,\mathbf{u}_{t})=\tilde{J}_{ET}(x_t,\mathbf{u}_{t})+\tilde{J}_{ER}(P_t,\mathbf{u}_{t})
=\sum_{k=0}^{N-1}\left[||x_{t+k}-p^s||_Q^2+||u_{t+k}||_R^2\right] +[||x_{t+N}-p^s||_{S}^2]
 +\sum_{k=0}^{N-1}\text{\text{trace}}(QP_{t+k})+\text{trace}(SP_{t+N}).$
 
The recursive feasibility and convergence analysis  for (\ref{TO11})-(\ref{TO41}) are established in the Theorem 5 and Theorem 6 as follows.

{\bf Theorem 5}. If there is a feasible solution for the optimization problem (\ref{TO11})-(\ref{TO41}) at time $t$, then it is recursively feasible.\\
{\sc Proof} See Appendix B.

{\bf Theorem 6}. Assume that the optimization problem (\ref{TO11})-(\ref{TO41}) admits a solution at time $t$, and there exists a matrix $W$ that satisfies $P_{t}\preceq W$, we then have
\begin{equation}\label{Y13}
\lim_{t\rightarrow \infty}\mathbb{E}\{||x_{t}-\hat{p}_t^s||\}\leq \left[\frac{N}{\lambda_{\min}(Q)}+\frac{\lambda_{\max}(S)}{\lambda_{\min}^2(Q)}\right]\text{\text{trace}}(SW).\nonumber
\end{equation}
{\sc Proof} See Appendix C.

\subsection{Recursive feasibility and convergence of MS-DCEE}

The recursive feasibility and convergence of the auxiliary MS-DCEE is established under the assumption that $\mathbb{E}(\hat{p}_{t}^s)=\bar{p}_{t}^s=p^s$ for $\forall \ t \in \mathbb{N}$, which is unrealistic. Actually, with the finite length of data $I_t$, the mean of the estimation yielded by Bayesian estimation is random, not only changing from one run to another but also changing with time $t$ in each run. This section aims to cope with this randomness with the help of the boundedness property in Section 4.1.

  In what follows, we will use the results of recursive feasibility and convergence of the auxiliary MS-DCEE to establish the recursive feasibility and convergence of the MS-DCEE by taking into account the transient Bayesian estimation error. To be specific, let us consider the real case where the mean of $\hat{p}_{t}^s$ asymptotically converges to the true source location $p^s$, i.e., $\mathbb{E}(\hat{p}_{t}^s)=\bar{p}_{t}^s\rightarrow p^s$ as $t\rightarrow\infty$. The recursively  feasibility
of (\ref{To1})-(\ref{To4}) is established by Theorem 7.

{\bf Theorem 7} Suppose that Assumption 1 holds.
If there is a feasible solution at time $t$, then the optimization problem (\ref{To1})-(\ref{To4}) with the MS-DCEE is recursive feasible.\\
{\sc Proof} See Appendix D.

The following lemma \citep{L1} plays an important role on development of the MS-DCEE.

{\bf Lemma 8}.
If $f(\cdot)$ is a continuous function, then there exists a $\mathcal{K}_\infty$ function $\alpha(\cdot)$ such that $|f(x)-f(y)|\leq \alpha(|x-y|)$, for all $x\in C$ and $y\in D$ with $C \subseteq D\subseteq \mathbb{R}^n$.

The main result is then presented by the following theorem.

{\bf Theorem 9} \label{convergence} Under assumption 1 and assumption  2, the closed loop system
(1) under the proposed MS-DCEE scheme is recursive feasible and satisfies the following conditions
    $$\lim_{t\rightarrow \infty}\mathbb{E}\{||x_{t}-\hat{p}_t^s||\}\leq \left[\frac{N}{\lambda_{\min}(Q)}+\frac{\lambda_{\max}(S)}{\lambda_{\min}^2(Q)}\right]\text{\text{trace}}(SW).$$

{\sc Proof} See Appendix E.

Theorem~\ref{convergence} states that with the carefully designed terminal cost and terminal constraint, it is guaranteed that the proposed MS-DCEE drives the agent to reach the source location with the error bounded.

\section{Simulations studies}
 To demonstrate the performance of the  proposed MS-DCEE algorithm, we compare the method with SMPC, IPP based on entropy proposed in \cite{Y1}, and the original single-step DCEE (without convergence guarantee) in \cite{C1} for searching the source of atmospheric hazardous material in a numerical simulation experiment. For the sake of fair comparison, we keep everything unchanged as in \cite{C1} except for the cost function and the terminal constraint as necessary in our scheme.
\begin{figure}
\centering
\includegraphics*[width=10cm]{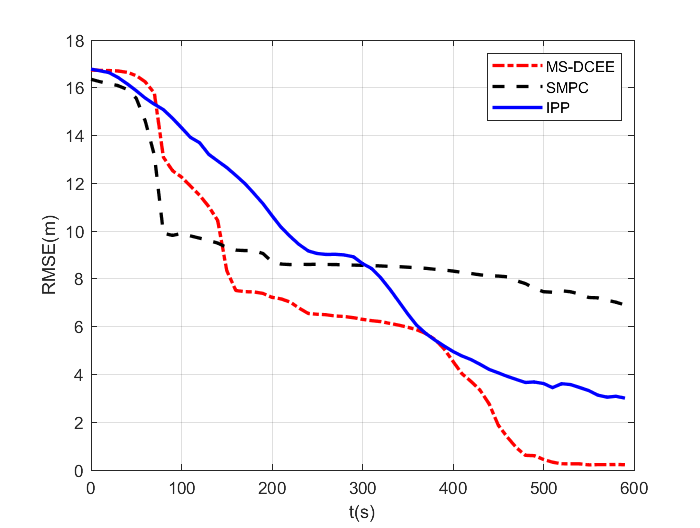}
\caption{The average root mean square error (RMSE) for estimated position and true source in the 120 runs.}\label{fig22}
\end{figure}

\begin{figure*}[htbp]
\centering
\includegraphics*[width=18cm]{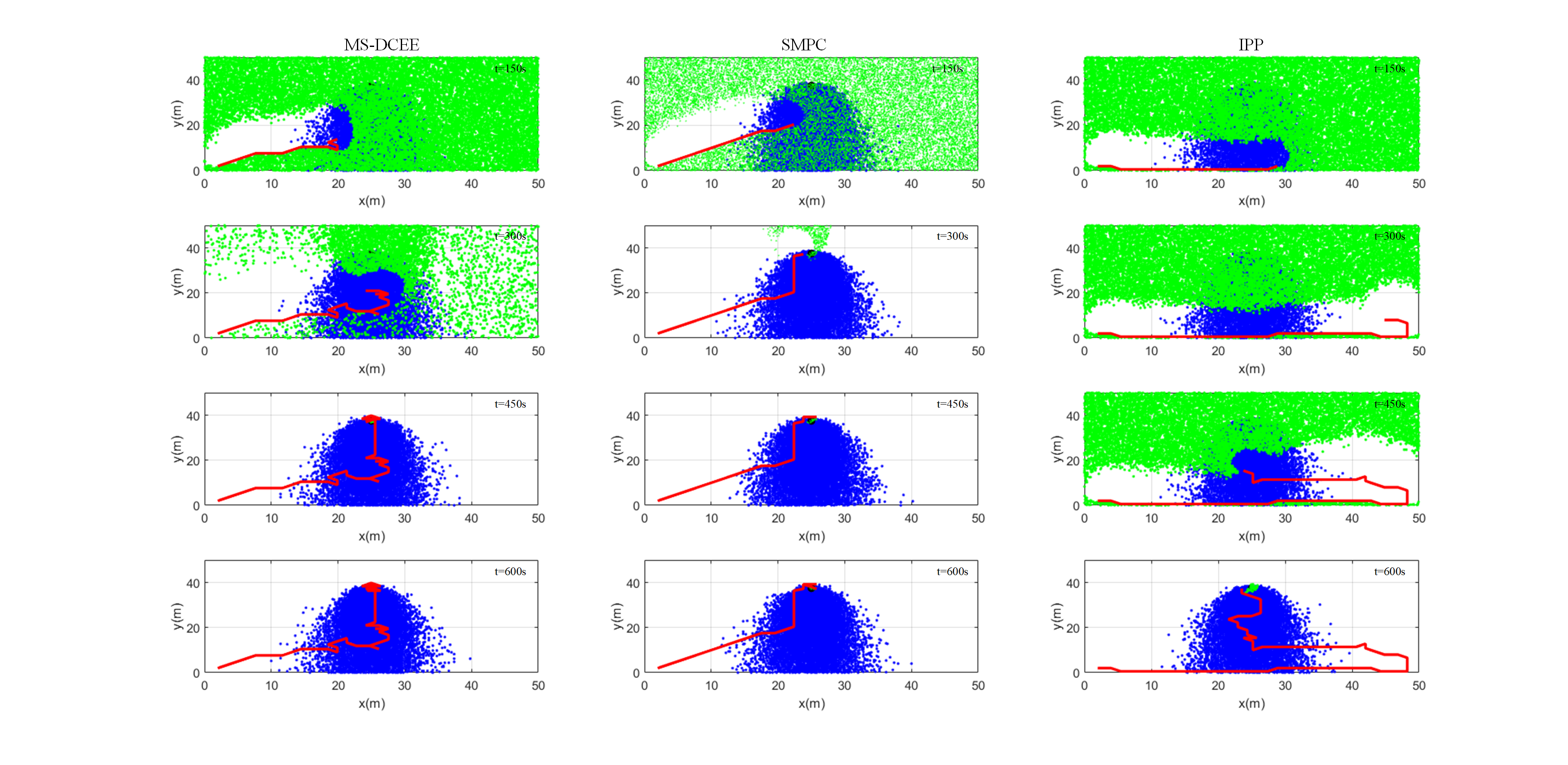}
\caption{Result of run for different algorithms on a single run at $150s, 300s, 450s$ and $600s$. Each small green dot represents a single
hypothesis of the source location. The red dot represents the
UAV current location and the red line represents the historic trajectory
taken.}\label{fig8}
\end{figure*}
In the simulation, we suppose $p_{z,t}=1$, $q_s=5g/s$, $u_s=4m/s$, $\phi_s=\frac{\pi}{2}$ $\zeta_1=1$ and $\zeta_2=8$ in (\ref{T303}). In order to reduce the amount of calculation, this paper selects the optimal control input from a limited set of actions. That is, $u_t\in \mathcal{U}:=\{\uparrow,\downarrow,\leftarrow,\rightarrow,\nwarrow,\nearrow,\swarrow,\searrow\}$ with a fixed step size of 2 m.  Sensor measurement noise is assumed to subject to Gaussian white noises with mean zero and standard deviation equal to 10\% of the signal. In order to reduce the randomness of detection, 4 different position of harmful atmospheric release with $[p^{s,x}=12.5, p^{s,y}=37.5, p^{s,z}=0]$, $[p^{s,x}=25.5, p^{s,y}=37.5, p^{s,z}=0]$, $[p^{s,x}=37.5, p^{s,y}=37.5, p^{s,z}=0]$ and $[p^{s,x}=25, p^{s,y}=10, p^{s,z}=0]$ are conducted.
\subsection{Performance comparisons between three algorithms}
The objective of this simulation study is to design an algorithm to quickly find pollution sources, we compare the performance of the proposed MS-DCEE algorithm, with the SMPC algorithm and the IPP algorithm.  The related  weights $Q=100I_2$, $R=100I_2$ and $S=161.8I_2$ are selected for SMPC and MS-DCEE algorithms. A Monte-Carlo simulation of 30 runs of every algorithm for four source configurations is conducted. After 120 Monte Carlo simulations for each search method, the result of the average root mean square error (RMSE) for the estimated position and the true source is shown in Fig. \ref{fig22}.

It can be seen from Fig. \ref{fig22} that the RMSE of the estimated source is in a downward trend, which is caused by the bayesian filter with new sample data. At the beginning, the root-mean-square descending gradient of the SMPC algorithm was significantly faster than the other two methods, which is caused by the initial configuration of the source, i.e., the robot moves toward the center of the search area near the true source. Around $120s$, the estimation accuracy of the SMPC algorithm began to slow down, because it did not learn future information. The MS-DCEE algorithm not only reduces the distance between the robot and the estimated mean value, but also learns the future uncertain information of the environment, which makes the accuracy of the estimation greatly improved.
At $490s$, the RMSE of MS-DCEE algorithm converges to a region close to 0. However, IPP algorithm will take more time to converge.

\begin{figure}
\centering
\includegraphics*[width=10cm]{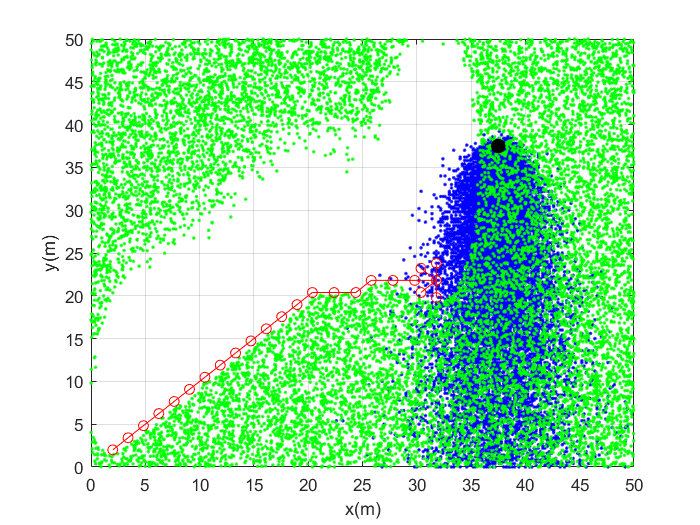}
\caption{Exemplar behaviour of SMPC in failed search cases.}\label{fig221}
\end{figure}

The results of the three methods at different time are shown in Fig. \ref{fig8}. The three columns in the figure represent MS-DCEE algorithm, SMPC algorithm and IPP algorithm, respectively. The four rows are results at time instants $t=150s, 300s, 450s, 600s$, respectively. In Fig. \ref{fig8}, the red line is the trajectory of the robot's movement, the blue dot indicates the concentration of chemical substances diffused (the denser the dots are, the higher the concentration that can be detected), and the green represents the particles used to estimate the source location.  It can be observed from the Fig. \ref{fig8} that the SMPC method takes the shortest time to search for and track the estimated source location, followed by the MS-DCEE, and time for IPP is the longest. The SMPC algorithm takes the shortest time because the source is close to the center of the search area. However, when the source is far from the center, as shown in Fig. \ref{fig221}, the SMPC algorithm can't find the source. 
\subsection{Performance of MS-DCEE algorithm with different predictive step}
In this subsection, to prove that the more information collected in the future, the better the optimized trajectory will be. By doing so, we compare the results of MS-DCEE with different prediction step $N$ ($N=1, 10, 20$).

\begin{figure}
\centering
\includegraphics*[width=10cm]{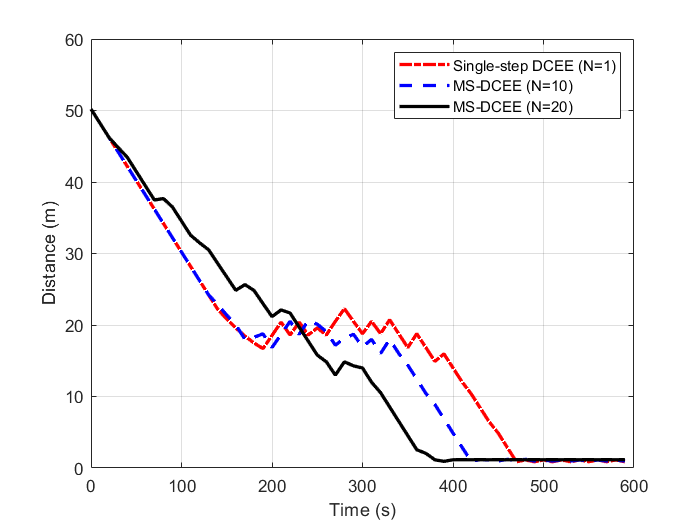}
\caption{Result of the MS-DCEE algorithm with different prediction step $N$.}\label{fig23}
\end{figure}

The results of MS-DCEE algorithm with three different prediction steps shown in Fig. \ref{fig23}. It can be seen that the time for the robot to reach the real source position are respectively, $t=390s$ for $N=20$, $t=420s$ for $N=10$, and $t=460s$ for $N=1$ (i.e., single-step DCEE algorithm). Therefore, we can conclude that the more information collected in the future, the less time it takes to find the source location.

\section{CONCLUSION}
A complete MS-DCEE framework with convergence guarantee is proposed to solve the problem of searching the source of airborne release. The MS-DCEE controller can actively probe and learn from the environment to reduce the uncertainty in its estimation of the source information and the environment. By balancing exploitation and exploration, the autonomous search algorithm strikes to reduce the variance of the source/environment estimation and drive the agent towards the believed source location. More importantly, the recursive feasibility and convergence of the proposed DCEE algorithm after modifying the original algorithm based on the conditions proposed in this paper are established.  Furthermore, it has been demonstrated that our algorithm achieves a better trade-off between SMPC algorithm and IPP algorithm and the benefits of using multiple steps prediction. Recently, it is noticed that the proposed DCEE actually share the same philosophy as Active Inference in neuroscience for interpreting perception, action and learning of human and animal intelligence \citep{action}. One future research direction is to build up understanding and cross fertilize between these two approaches.

{\bf Acknowledgem}

This work was supported by the UK Engineering and Physical Sciences Research Council (EPSRC) Established Career Fellowship ``Goal-Oriented Control Systems: Disturbance, Uncertainty and Constarints'' under the grant number EP/T005734/1.

{\bf Appendix}
\begin{appendix}
\section{Proof of Theorem 4}
According to the result (i) of Lemma 3, we get $\mathbb{E}(||\hat{p}_{t}^s-p^s||^2)\leq \text{trace}(W)$. One further has $\lim_{t\rightarrow\infty}\mathbb{E}\left(\hat{p}_{t}^s\right)=p^s$ following (ii) of Lemma 3. As such, for $t\rightarrow\infty$, one obtains
\begin{eqnarray}
\mathbb{E}(||\hat{p}_{t}^s-p^s||^2)&=&\mathbb{E}\left(||\hat{p}_{t}^s-\mathbb{E}(\hat{p}_{t}^s)+\mathbb{E}(\hat{p}_{t}^s)-p^s||^2\right)\nonumber\\
&=&\mathbb{E}(||\hat{p}_{t}^s-\bar{p}_{t}^s||^2)=\text{trace}(P_{t})
\end{eqnarray}

\section{Proof of Theorem 5}
At time instant $t$, assume the optimal solution  given by the control sequence $\mathbf{u}_{t}^*=[u_{t}^*,u_{t+1}^*,\cdots,u_{t+N-1}^*]$ with the predicted state trajectory  $\mathbf{x}_{t}^*=[x_{t},x_{t+1}^*,x_{t+2}^*,$ $\cdots,x_{t+N-1}^*,x_{t+N}^*]$ satisfying $x_{t+N}^*-p^s\in \tilde{\mathcal{T}}$, i.e., $x_{t+N}^*\in \{p^s\}\oplus \tilde{\mathcal{T}}\subseteq\mathcal{X}$. At the next time step $t+1$, the sub-optimal control  sequence and the corresponding state trajectory are denoted by $\mathbf{u}_{t+1}=[u_{t+1}^*,u_{t+2}^*,\cdots,u_{t+N-1}^*, u_{t+N}]$ and $[x_{t+1}^*,x_{t+2}^*,\cdots,x_{t+N}^*,$ $x_{t+N+1}]$ with $u_{t+N}=K(x_{t+N}^*-p^s)\in \mathcal{U}$ and $x_{t+N+1}-p^s=x_{t+N}^*-p^s+u_{t+N}$  satisfying $x_{t+N+1}-p^s=(I_3+K)(x_{t+N}^*-p^s)\in \tilde{\mathcal{T}}$ .  The state $x_{t+N+1}\in \{p^s\}\oplus\tilde{\mathcal{T}}\subseteq\mathcal{X}$ is trivially satisfied.

\section{Proof of Theorem 6}
For simplicity of notation, we refer to $\tilde{J}(t), \tilde{J}_{ET}(t)$, $\tilde{J}_{ER}(t)$ as $\tilde{J}(x_t,\mathbf{u}_{t}), \tilde{J}_{ET}(x_t,\mathbf{u}_{t}), \tilde{J}_{ER}(x_t,\mathbf{u}_{t})$. This proof is mainly conducted via following three steps.

\textit{Step 1}. From the definition of $\tilde{J}(t)$, we obtain the lower bound of $\tilde{J}^*(t)$:
\begin{align}\label{Y6}
&\tilde{J}^*(t)\geq\mathbb{E}\{||x_{t}-\hat{p}_t^s||_Q^2\}+||u_t||_R^2\nonumber\\
&\geq\lambda_{\min}(Q)\mathbb{E}\{||x_{t}-\hat{p}_t^s||^2\}
\end{align}
where $\lambda_{\min}(Q)$ is the smallest eigenvalue of the positive matrix $Q$.

\textit{Step 2}. We define a feasible control sequence ${\bf u}_t=\left[K(x_t-p^s),\cdots,K(I_3+K)^{N-1}(x_t-p^s)\right]$ at time $t$. Recall (\ref{Y1}), and we have
\begin{equation}
~~~~~~\tilde{J}_{ET}(t)=||x_t-p^s||_S^2.
\end{equation}
Since $S\succeq Q$, $P_{t+N}\preceq W$, one obtains
\begin{equation}
\tilde{J}_{ER}(t)\leq \text{trace}(SP_{t})+N\cdot\text{trace}(SW).
\end{equation}
In view of the optimality of $\tilde{J}^*(t)$,  we have
\begin{eqnarray}\label{Y10}
\tilde{J}^*(t)\leq  \tilde{J}(t)=\tilde{J}_{ET}(t)+\tilde{J}_{ER}(t)\leq \lambda_{\max}(S)\mathbb{E}\{||x_{t}-\hat{p}_t^s||^2\}+N\cdot\text{trace}(SW)
\end{eqnarray}
where $\lambda_{\max}(S)$ is the biggest eigenvalue of the positive matrix $S$.
This proves the upper bound of $\tilde{J}^*(t)$.

\textit{Step 3}. The relationship between $\tilde{J}^*(t)$ and $\tilde{J}^*(t+1)$ is established.
Now assume that the optimal cost function with the optimal control sequence $\mathbf{u}_{t}^*$ at time $t$ is $J^*(t)$. At the next time $t+1$, a suboptimal control sequence is denoted by $\mathbf{u}_{t+1}=[u_{t+1}^*,u_{t+2}^*,\cdots,u_{t+N-1}^*,u_{t+N}]$  with $u_{t+N}=K(x_{t+N}^*-p^s)$, it thus follows that the sub-optimal cost function  $\tilde{J}(t+1)$  satisfies
\begin{equation}\label{Y2}
\ \ \ \ \  \ \ \ \ \tilde{J}^*(t+1)\leq \tilde{J}(t+1).
\end{equation}
In view of the definition of $S$ given by (\ref{Y1}), $\tilde{J}(t+1)$ is equivalent to the following equation
\begin{eqnarray}\label{Y3}
\tilde{J}(t+1) = \tilde{J}^*(t)-||x_{t}-p^s||_Q^2-||u_{t}^*||_R^2-\text{trace}(QP_t) -\text{trace}(SP_{t+N})+\text{trace}(QP_{t+N})+\text{trace}(SP_{t+N+1}).\nonumber
\end{eqnarray}
Since $S\succeq Q$, $P_{t+N+1}\preceq W$, one has
\begin{eqnarray}\label{Y5}
\!\!\!\!\!\!\!\!\!\!\tilde{J}^*(t+1)\leq \tilde{J}^*(t)-\lambda_{\min}(Q)\mathbb{E}\{||x_{t}-\hat{p}_t^s||^2\}+\text{trace}(SW).
\end{eqnarray}
It further follows from
 (\ref{Y10}) that
\begin{eqnarray}\label{Y100}
-\mathbb{E}\{||x_{t}-\hat{p}_t^s||^2\}\leq \frac{N\cdot \text{trace}(SW)-\tilde{J}^*(t)}{\lambda_{\max}(S)}.
\end{eqnarray}
Combining (\ref{Y5}) and (\ref{Y100}), one obtains
\begin{eqnarray}\label{Y11}
\tilde{J}^*(t+1)\leq \tilde{J}^*(t)\left[1-\frac{\lambda_{\min}(Q)}{\lambda_{\max}(S)}\right]+\left[\frac{\lambda_{\min}(Q)}{\lambda_{\max}(S)}N+1\right]\text{trace}(SW).\nonumber
\end{eqnarray}
 We now denote a compact set $$\Omega=\left\{x| \tilde{J}(t)\leq\left[N+\frac{\lambda_{\max}(S)}{\lambda_{\min}(Q)}\right]\text{trace}(SW)\right\}.$$
If $x_t\in \mathcal{X} \setminus \Omega$, then we have $$\tilde{J}^*(t)\frac{\lambda_{\min}(Q)}{\lambda_{\max}(S)}>\left[\frac{\lambda_{\min}(Q)}{\lambda_{\max}(S)}N+1\right]\text{trace}(SW), $$
which implies that $\tilde{J}^*(t+1)<\tilde{J}^*(t)$ and $\tilde{J}^*(t)$ is decreasing from the beginning until
$x_t$ enters $\Omega$.
Once $x_t$ enters $\Omega$, in view of (\ref{Y11}), we have
\begin{equation}\label{Y223}
\tilde{J}^*(t+1)\leq\left[N+\frac{\lambda_{\max}(S)}{\lambda_{\min}(Q)}\right]\text{\text{trace}}(SW),
\end{equation}
which implies that once the state $x_t$ enters $\Omega$, its successor $x_{t+1}$ will definitely in $\Omega$ as well. This indicates that with the proposed MS-DCEE, the state $x_t$ will enter the compact set $\Omega$ in finite time and remains therein.

Combining the lower bound of $\tilde{J}^*(t)$ in (\ref{Y6}), it holds that
\begin{equation}\label{Y13}
\lim_{t\rightarrow \infty}\mathbb{E}\{||x_{t}-\hat{p}_t^s||\}\leq \left[\frac{N}{\lambda_{\min}(Q)}+\frac{\lambda_{\max}(S)}{\lambda_{\min}^2(Q)}\right]\text{\text{trace}}(SW).\nonumber
\end{equation}

\section{Proof of Theorem 7}
Slightly different from the proof of Theorem 5, the last term of the sub-optimal control sequence and state sequence are now
 $u_{t+N}=K(x_{t+N}^*-\bar{p}_{t+N}^s)+K_f\delta_{t+N}$ and $x_{t+N+1}=x_{t+N}^*+u_{t+N}$ at time step $t+1$ with $K_f=I_3$. Since $\delta_{t+N}=\bar{p}_{t+N+1}^s-\bar{p}_{t+N}^s$, we have
\begin{align}
x_{t+N+1}=x_{t+N}^*+K(x_{t+N}^*-\bar{p}_{t+N}^s)+\delta_{t+N}.
\end{align}
Therefore, one has $x_{t+N+1}-\bar{p}_{t+N+1}^s=(I_3+K)(x_{t+N}^*-\bar{p}_{t+N}^s)$ and $x_{t+N}^*-\bar{p}_{t+N}^s\in \mathcal{T}$, which implies that $x_{t+N+1}-\bar{p}_{t+N+1}^s=(I_3+K)(x_{t+N}^*-\bar{p}_{t+N}^s)\in \mathcal{T}$. According to Assumption 5, it can be verified that $u_{t+N}=K(x_{t+N}^*-\bar{p}_{t+N}^s)+\delta_{t+N}\in (K\circ \mathcal{T})\oplus(K_f\circ\mathcal{G})\subseteq \mathcal{U}$, $x_{t+N+1}=(I_3+K)(x_{t+N}^*-\bar{p}_{t+N}^s)+\bar{p}_{t+N+1}^s\in \mathcal{T}\oplus\mathcal{O}\subseteq \mathcal{X}$.
\section{Proof of Theorem 9}
 Theorem 4 indicates that the mean of $\hat{p}_{t}^s$ asymptotically converges to the true source location $p^s$, i.e., $\mathbb{E}(\hat{p}_{t}^s)=\bar{p}_{t}^s=p^s$ for $t\rightarrow\infty$. Since $J(t)$ in (\ref{TY23}) and $\tilde{J}^*(t)$  in (\ref{TO11}) have the same structure and are both continuous function, it follows from Lemma 8 that there exists a $\mathcal{K}_\infty$ function $\alpha(\cdot)$ such that
\begin{equation}
~~~~~~~J^*(t)-\tilde{J}^*(t)\leq \alpha(|\bar{p}_t^s-p^s|),
\end{equation}
for all $x_t\in \mathcal{X}, \bar{p}_t^s\in \mathcal{X}$ and $p^s\in \mathcal{X}$.
Therefore,  for $t\rightarrow\infty$, $\alpha(0)=0$, we have
$$\lim_{t\rightarrow \infty}\mathbb{E}\{||x_{t}-\hat{p}_t^s||\}\leq \left[\frac{N}{\lambda_{\min}(Q)}+\frac{\lambda_{\max}(S)}{\lambda_{\min}^2(Q)}\right]\text{\text{trace}}(SW).
$$
\end{appendix}

\end{document}